\documentclass[aps,preprint,nofootinbib,superscriptaddress]{revtex4}
\usepackage{graphicx,color}
\usepackage{amsmath,amssymb,amscd,amsfonts}
\usepackage[caption=false]{subfig}
\usepackage{slashed}
\usepackage{bm}
\usepackage{xcolor}
\usepackage{diagbox,pict2e}

\def \Act{\mathcal A}
\newcommand{\bwt}{\begin{widetext}}
\newcommand{\ewt}{\end{widetext}}
\newcommand{\newc}{\newcommand}
\newc{\hc}{\dagger}
\newc{\pd}{\partial}
\newc{\beq}{\begin{equation}}
\newc{\eeq}{\end{equation}}
\newc{\beqa}{\begin{eqnarray}}
\newc{\eeqa}{\end{eqnarray}}
\newc{\nonr}{\nonumber}
\newc{\bi}{\begin{itemize}}
\newc{\ei}{\end{itemize}}
\newc{\ra}{\rightarrow}
\newc{\la}{\leftarrow}
\newc{\lra}{\longrightarrow}
\newc{\lla}{\longleftarrow}
\newc{\Lra}{\Longrightarrow}
\newc{\Lla}{\Longleftarrow}
\newc{\half}{\frac{1}{2}}
\newc{\del}{\delta}
\newc{\Del}{\Delta}
\newc{\eps}{\epsilon}
\newc{\gm}{\gamma}
\newc{\lam}{\lambda}
\newc{\kap}{\kappa}
\newc{\tri}{\triangle}
\newc{\epsp}{\epsilon^\prime}
\newc{\wt}{\widetilde}
\newc{\ovl}{\overline}
\newc{\tchi}{\tilde{\chi}}
\newc{\ds}{\displaystyle}
\newc{\pmt}{\pm\!\pm}
\newc{\PL}{\hat{L}}
\newc{\PR}{\hat{R}}
\newc{\st}{s_\theta}
\newc{\ct}{c_\theta}

\newcommand{\Uel}{U(1)_\ell}
\newc{\msm}{\mathrm{SM}}
\newc{\mtev}{\mathrm{TeV}}
\newc{\mgev}{\mathrm{GeV}}
\newc{\Tr}{\mathrm{Tr}}
\newc{\clbl}{\color{blue}}
\newc{\clg}{\color{green}}
\newc{\clr}{\color{red}}

\mathchardef\mhyphen="2D
\newc{\SL}{\not\!\!}

\begin{document}
\title{Alternative Perspective on Gauged Lepton Number and Implications for Collider Physics}
\date{\today}

\author{We-Fu Chang}
\email{wfchang@phys.nthu.edu.tw}
\affiliation{Department of Physics, National Tsing Hua University, No. 101, Section 2, Kuang-Fu Road, Hsinchu, 30013,  Taiwan}
\affiliation{TRIUMF, 4004 Wesbrook Mall, Vancouver, BC, V6T 2A3, Canada}
\author{John N. Ng}
\email{misery@triumf.ca}
\affiliation{TRIUMF, 4004 Wesbrook Mall, Vancouver, BC, V6T 2A3, Canada}

\begin{abstract}
A new anomaly-free gauged $U(1)_\ell$ lepton-number model is studied. Two standard model lepton generations acquire the same but oppositive sign $U(1)_\ell$ charges, while four exotic chiral leptons cancel the anomalies of the remaining lepton family.
We discuss a simplified case which has the universal Yukawa couplings. It agrees with all the experimental constraints and predicts $m_e, m_\mu \ll m_\tau$, and the latter is of the electroweak scale. Due to the interference between the SM and $U(1)_\ell$ gauge interactions, this model robustly predicts that $e,\mu,\tau$ have distinctive
forward-backward asymmetries at the $e^+e^-$ colliders. It can be searched for at the $e^+e^-$ machine with $\sim$ TeV center-of-mass energy and an integrated luminosity $\sim ab^{-1}$.

\end{abstract}

\maketitle
\section{Introduction}

The Standard Model (SM) of particle physics based on the gauge group $SU(3)\times SU(2)\times U(1)$ is spectacularly successful in explaining current data. It contains two accidental symmetries associated with
lepton- and baryon-number conservation. The structure of the model cannot explain their occurrence. Furthermore, the minimal version cannot accommodate neutrino masses
which are indicated by neutrino oscillation data. Without adding any new degrees of freedom, finite
neutrino masses can be induced by adding a dimension-five  Weinberg operator \cite{WO}, $O_5= \frac{c}{\Lambda} \ell_L \ell_L H H$. Where $\ell_L$ denotes the SM lefthanded doublet, $H$ is the SM Higgs field,  $\Lambda $ is an unknown cutoff scale, and $c$ is a free parameter. After $H$ takes a vacuum expectation value $v\simeq 247 \mgev$, a neutrino mass $m_\nu \sim \frac{cv^2}{\Lambda}$ is generated. This operator
breaks the lepton-number. In order to satisfy the experimental limit of $m_\nu \lesssim 1 \mathrm{eV}$,
the scale $\Lambda$ must be in the range of 1 to $10^{11} \mtev$.
This path for neutrino masses generation indicates that the SM is an effective theory and it
has to be extended.

From the discussion above, it is clear that neutrino masses and the nature of lepton-number are closely
related. With the usual lepton-number $\ell$ assignments, i.e. the charged leptons $e,\mu,\tau$ and
their neutrino partners have $\ell=1$, and anti-leptons have $\ell=-1$, $O_5$ breaks $\ell$ by two units.
Moreover, whether the lepton-number symmetry, taken to be $U(1)_\ell$, is a global or local gauge symmetry is left unanswered. If $U(1)_\ell$ were a broken global symmetry, a massless Goldstone boson, the Majoron, will be generated \cite{CMP}. The astroparticle and cosmological consequences of this case was studied in \cite{CNM1,CNM2,CNM3}.
On the other hand, for a broken gauged $U(1)_\ell$ the Goldstone boson will become the longitudinal component of a massive leptophilic gauge boson $Z_\ell$. The existence of $Z_\ell$ is a robust prediction
if lepton-number is a broken gauged Abelian symmetry. It is also well-known that the SM is anomalous under $U(1)_\ell$. How one solves these anomalies requires a more in-depth look into the nature of lepton-number.

Historically, the three SM lepton generations are given different names or quantum numbers, $e,\mu,\tau$, and
were taken to be conserved\footnote{
Different conserved electron and muon quantum numbers were first introduced in \cite{FW} to explain the non-observation of $\mu\ra e \gm$ for massless neutrinos. Currently, within the SM this decay has a tiny branching ratio $\lesssim 10^{-45}$ due to the neutrino masses $\lesssim 1$ eV; thus, eliminating the
 need for these conserved quantum numbers. In addition, all SM processes measured are not sensitive to what these quantum numbers are.}. With the discovery of neutrino oscillations these quantum numbers can no longer be conserved.
Nevertheless; they serve as efficient bookkeeping devices. In most studies, they all are assigned with the same lepton-number $\ell=1$. In this paper, we shall refer to them as first, second and third generations, and reserve the flavor labeling $e,\mu,\tau$ to denote  the charged lepton in the mass basis with eigenvalues $m_e, m_\mu, m_\tau$, respectively.
The anomalies can be associated with total lepton-number, and new leptons are added to cancel this
total lepton-number anomaly as is done in \cite{STV}. If one assigns the same value $\ell=1$ to all the
SM leptons, as is done conventionally, the corresponding anomaly can also be solved for each generation \cite{CNL1,CNL2}. Both solutions involve many extra new leptons.

In this paper, we point out that setting $\ell=1$ for all SM leptons is not necessary for a gauged $U(1)_\ell$, and the three generations can have different lepton charges and one universal gauge
couplings $g_\ell$. This simple observation amounts to taking $U(1)_\ell$ to be entirely analogous to QED
where different particles can have the different amount of charges but one universal coupling, $e$.
Specifically, we can have $\ell=\zeta_1, \zeta_2,\zeta_3$ for the three SM lepton generations in no particular order,
and $\zeta_1, \zeta_2,\zeta_3$ are real numbers. In general, they can all be different. The case in which $\zeta_1= \zeta_2=\zeta_3=1$ was extensively studied in \cite{STV,CNL1,CNL2}and earlier references there.

 We further note that if two generations have equal and opposite lepton charges, e.g., $\zeta_2=-\zeta_1$, the anomalies cancelation can be achieved with a significantly reduced number of new leptons required. This is easy to see since we only need {\em one set} of vectorlike new leptons for a single generation anomalies cancelation if that remaining generation has nonvanishing lepton charge, i.e., $\zeta_3\neq 0$\footnote{If $\zeta_3=0$ this is the same as having conserved $\ell_\mu-\ell_\tau$ \cite{HJLV,Fo}.}. Details of the quantum number assignments will be given in Sec. 2. In Sec. 3 details the charged lepton mass matrix and its diagonalization are given. This is a non tirvial issue since the new charged leptons can mix with the SM partners. In Sec.4 we carefully study the SM gauge interactions with the presence of the exotic leptons which carry the SM quantum numbers.
Experimental constraints on the mixings among the exotic leptons and the SM ones have to be carefully implemented. The resulting phenomenology of the $Z_\ell$ has interesting features that distinguish from previous studies. This is given in Sec. 5.  Sec. 6 contains our conclusions. In general one can have
kinetic mixing between $U(1)_Y$ and $\Uel$ \cite{Holdom}, which is expected to be small. The
phenomenology of this mixing was discussed in detail in \cite{CNW06} and references therein. These considerations will not be repeated here.

\section{Anomalies cancelation for $\Uel$}
We extend the SM gauged group by adding a $\Uel$ and is explicitly given as $G=SU(2)\times U(1)_Y\times\Uel$.
First, discuss the anomalies of a single family.
 We assume that both the left-handed and right-handed SM leptons carry $\Uel$ charge $\zeta$.

The new anomaly coefficients are
% \begin{subequations}
 \beqa
&& \Act_1([SU(2)]^2\Uel)=-\zeta /2\,,\, \Act_2([U(1)_Y]^2\Uel)=\zeta/2\,, \nonr\\
&& \Act_3([U(1)_Y[\Uel]^2)=0\,,\, \Act_4([\Uel]^3)=-\zeta^3\,,\,  \Act_5(\Uel)=-\zeta\,,
  \label{eq:ac}
 \eeqa
% \end{subequations}
 where $\Act_5$ stands for the lepton-graviton anomaly.
 While new chiral leptons are introduced to cancel Eq.(\ref{eq:ac}), one also needs to make sure that the SM anomalies of $\Act_6([SU(2)]^2 U(1)_Y)$, $\Act_7([U(1)_Y]^3)$,  and $\Act_8(U(1)_Y)$ are canceled.
  It is easy to check that the new vectorlike leptons in Table.\ref{tb:lA} cancel the above anomalies.
 %\vspace{1cm}
\begin{table}
\begin{center}
\renewcommand{\arraystretch}{1.30}
\begin{tabular}{|c| c c ccc c c c|}
\hline
Field& $\ell_L=\left(\begin{smallmatrix}l^0_L\\ l^-_L\end{smallmatrix} \right)$
& $l_R$
& & $L_{1L}= \left(\begin{smallmatrix}N_{1L}\\ E_{1L}\end{smallmatrix}\right)$
& $E_{1R}$
& & $L_{2R}= \left(\begin{smallmatrix}N_{2R}\\ E_{2R}\end{smallmatrix} \right)$
& $E_{2L}$ \\ \hline\hline
$SU(2)$& $2$ & $1$ && $2$ & $1$ && $2$ & $1$ \\
$U(1)_Y$& $-\frac{1}{2}$ & $-1$ && $-\frac{1}{2}$ & $-1$&& $-\frac{1}{2}$ & $-1$ \\
$U(1)_\ell$& $\zeta$ & $\zeta$ &&$-\zeta$ &$-\zeta$ &&$0$ &$0$ \\
\hline
\end{tabular}
\caption{Lepton fields for anomaly-free solution.}
\label{tb:lA}
\end{center}
\end{table}
%\vspace{.5cm}
Since the pair of new leptons are vectorlike, the SM anomalies $\Act_6([SU(2)]^2 U(1)_Y)$, $\Act_7([U(1)_Y]^3)$, and $\Act_8(U(1)_Y)$ cancelations are not affected. %\footnote{It is easy to check that if $\ell=a$ for the SM leptons where $a$ is an integer the solution for the vectorlike leptons will be $\ell_1=a,\ell_2=0$.}.
These are the simplest solutions we found. If one allows the two new doublets to have hypercharge $Y> 1/2$, then all anomalies are canceled with the following set of vectorlike leptons:$L_{1L}:(2,\frac{7}{2},2\zeta); L_{2R}:(2,\frac{7}{2},3\zeta);E_{1R}:(1,5,3\zeta); E_{2L}: (1,5,4\zeta)$, where the notation follows that of Table(I). Since these states will have high electric charges and are stable, they are ruled out experimentally. Our solution is the only viable one with rational lepton charges.

It is also clear that if two generations have equal and opposite lepton charges, then each one of Eq.(\ref{eq:ac}) will exactly cancel between the two families. For this case, there is no need to introduce new fermions for anomaly cancelation \cite{HJLV,Fo}.
This gives a simple solution to the 3 generation case: arranging two generations to have equal and opposite $\ell$ values, say $\zeta_2=-\zeta_1$,  and the remaining generation is given by Table (I) which is anomaly free has $\ell=\zeta_3$.
For $|\zeta_1|\neq \zeta_3$, the three generations do not mix  and the Yukawa couplings are only allowed within each generation, so that the flavor basis coincides with the lepton-number basis.

Without loss of generality, we can normalize the lepton charge such that $\zeta_3=1$.
Then in general, $|\zeta_1|$ needs not be $\zeta_3(=1)$. However, in this paper, we are interested in $|\zeta_1|=1$ as it presents an interesting and  novel phenomenology. This is due to the essential  mixings of the SM leptons originating from the lepton charge assignments consistent with anomalies cancelation.
In this case, flavor labels will be meaningful only after charged lepton mass diagonalization.
We shall see in the later section that it might provide a partial understanding of  why $m_e,m_\mu\ll m_\tau$.
The case $|\zeta_1|\neq 1$ will be left for a future study.

It is convenient to use the following $(SU(2),U(1)_Y,\Uel)$ designations.
The SM leptons are denoted  as following: $l_{L1,L2}(2,-1/2,1)$, $l_{L3}(2,-1/2,-1)$, $e_{R1,R2}(1,-1,1)$, and $e_{R3}(1,-1,-1)$.
Also, for the exotic leptons, $l_{L4}\equiv L_{1L}(2,-1/2,-1)$, $e_{R4}\equiv E_{1R}(1,-1,-1)$.
$L_{2R}(2,-1/2,0)$ and $E_{2L}(1,-1,0)$ retain their names as in Table I.
We emphasize again that at this stage the generation indices have nothing to do with the lepton flavor yet. The lepton flavor appears only after the mass diagonalization.

\section{Charged Lepton Masses}

Besides the SM Higgs doublet, $H(2,1/2,0)$, a singlet scalar $\phi_1(1,0,1)$ is introduced
for $\Uel$ symmetry breaking, and
to make the exotic charged lepton heavier than the Fermi scale as in \cite{CNL1,CNL2}.
The $G$ invariant Yukawa interaction is
\beqa
&&\sum_{i,j=1,2}y_{ij}\,\bar{l}_{Li} H e_{Rj} + \sum_{a,b=3,4} y_{ab}\,\bar{l}_{La} H e_{Rb} +y_{55}\,\bar{L}_{2R} H E_{2L} \nonr\\
&&+ \sum_{i=1,2}( f_i\, \bar{l}_{Li} L_{2R}  + f_i^\prime \bar{e}_{Ri}E_{2L} ) \phi_1 + \sum_{a=3,4}( f_a\, \bar{l}_{La} L_{2R}
+f_a^\prime \bar{e}_{Ra}E_{2L} )\phi_1^* +H.c.
\eeqa

After $H$ and $\phi_1$ develop VEVs, $\langle H \rangle= \frac{v}{\sqrt{2}}\left(\begin{smallmatrix} 0\\ 1\end{smallmatrix} \right) $ and $\langle \phi_1\rangle=v_L/\sqrt{2}$, respectively, the charged lepton Dirac mass matrix in the basis of $\{e_1,e_2,e_3,e_4,E_2\}$ becomes\footnote{The intent to begin with a basis where the upper-left $4\times 4$ mass matrix is diagonal, i.e., $\eps_{2,3,6,7}=0$, does NOT help since this is not the mass eigenstate and this choice requires elaborated fine tuning  to reproduce the observed charged lepton masses.  }
\beq
\label{eq:CLM}
{\cal M}^c=\frac{v_L}{\sqrt{2}}\left(
  \begin{array}{ccccc}
 \eps_1& \eps_2&0&0&f_1\\
  \eps_3& \eps_4&0&0&f_2\\
  0&0& \eps_5& \eps_6 &f_3\\
   0&0& \eps_7& \eps_8 &f_4\\
   f_1^\prime & f_2^\prime &f_3^\prime &f_4^\prime& \eps_9
  \end{array}\right)\,.
\eeq
The same Yukawa interaction with $f_{1,2,3,4}$ also gives Dirac masses to the neutral leptons.
In the basis of $\{\nu_1, \nu_2, \nu_3, N_1, N^c_2\}$, the mass matrix is
\beq
\label{eq:NLM}
{\cal M}^n=\frac{v_L}{\sqrt{2}}\left(
  \begin{array}{ccccc}
 0& 0&0&0&f_1\\
  0& 0&0&0&f_2\\
  0&0& 0& 0&f_3\\
   0&0& 0& 0&f_4\\
   f_1 & f_2 &f_3 &f_4& 0
  \end{array}\right)\,.
\eeq

Without tuning, we expect $f_i, f'_i \sim {\cal O}(1)$, $\eps_i\sim {\cal O}(v/v_L)$, and the Yukawa couplings are not displayed. In general, ${\cal M}^c$ is not symmetric but it can be
 diagonalized by a bi-unitary rotation such that $(U_L)^\dagger\cdot {\cal M}^c\cdot U_R=\mbox{diag}(m_e,m_\mu,m_\tau, M_-, M_+)$. To proceed, we need further assumptions on the various Yukawa couplings.

It is instructive to consider the limiting  case of  $f_i=f'_i =1, \eps_i=\eps  \forall\, i $
 which will be referred to as equal Yukawa limit (EYL).
This will give a symmetric mass matrix with two zero eigenvalues which contradicts the experimental facts that $m_e,m_\mu\neq 0$ but $m_e,m_\mu\ll v $. In order to generate these two
 small values (for $e$ and $\mu$), the perturbations $\delta_1<\delta_2 \ll 1$ are introduced. Thus,
\beq
\label{eq:SCLM}
{\cal M}^{\prime c}=\frac{v_L}{\sqrt{2}}\left(\begin{array}{ccccc}
 \eps& \eps(1-\delta_1)&0&0&1\\
  \eps(1-\delta_1)& \eps&0&0&1\\
  0&0& \eps& \eps(1-\delta_2) &1\\
   0&0& \eps(1-\delta_2)& \eps &1\\
   1 & 1 &1 &1& \eps
  \end{array}\right) \,,
\eeq
 and there is no change to ${\cal M}^n$.
The simplified mass matrix, ${\cal M}^{\prime c}$ can be diagonalized, to the leading order, by an orthogonal transformation,
\beq
\label{eq:SUR}
U=\left(
    \begin{array}{ccccc}
      \frac{1}{\sqrt{2}} & 0 & -\frac{1}{2} & -\frac{1}{2\sqrt{2}} & \frac{1}{2\sqrt{2}} \\
 -\frac{1}{\sqrt{2}} & 0 & -\frac{1}{2} & -\frac{1}{2\sqrt{2}} & \frac{1}{2\sqrt{2}} \\
0&  \frac{1}{\sqrt{2}}  & \frac{1}{2} & -\frac{1}{2\sqrt{2}} & \frac{1}{2\sqrt{2}} \\
0&  -\frac{1}{\sqrt{2}}  & \frac{1}{2} & -\frac{1}{2\sqrt{2}} & \frac{1}{2\sqrt{2}} \\
0&0&0&\frac{1}{\sqrt{2}} &\frac{1}{\sqrt{2}}
    \end{array}
  \right)\,,
\eeq
and $  U^T\cdot {\cal M}^{\prime c}\cdot U \simeq (v_L/\sqrt{2})\times\mbox{diag}\{\delta_1 \epsilon, \delta_2 \epsilon, 2\epsilon, -2+3\epsilon/2, 2+3\epsilon/2\}$.
 The neutral lepton mass matrix is diagonalized by the very same rotation, $  U^T\cdot {\cal M}^{n}\cdot U \simeq (v_L/\sqrt{2})\times\mbox{diag}\{0, 0, 0, -2, 2\}$, namely the light and heavy neutrinos decouple at the leading order.
In the mass basis, there is a heavy Dirac neutrino pair with a mass at the lepton number breaking scale, and  the three light SM neutrinos are massless.
The realistic neutrino masses need further model building, see the remark in the conclusion section.

 This limiting case provides an interesting feature that two out of three SM charged leptons are below the Fermi scale, one is at electroweak scale, and two at the lepton symmetry scale.  Notice that this mass hierarchy  does not require tuning Yukawa's but come from a more symmetric structure
and the above statement holds in the leading approximation. It is natural to identify the first two light states to be the $e,\mu$, the third one as the $\tau$, and the two heavy ones as new yet to be discovered leptons with masses at the lepton symmetry breaking scale. Thus we recovered the SM flavor structure. Each physical state, except $e,\mu$,  is a linear combination of at least four gauge states.  This mechanism  is reminiscent of Type I seesaw neutrino mass generation. The two heavy
leptons $E_{1,2}$ play the similar role of heavy sterile neutrinos in the seesaw case. Here they
arise naturally from anomalies cancelation and not put in by hand.

 Now, the universal Yukawa coupling, $y$, in this EYL can be fixed by tau mass, $ y= \sqrt{2} m_\tau/v$. Moreover, the splitting parameters are
determined to be $\delta_1=m_e/m_\tau$ and $\delta_2=m_\mu/m_\tau$ as well.
The Higgs couplings in this simple EYL scenario reproduce the general feature of the Higgs portal models.
In the mass basis, the couplings of the three light charged leptons to the 125 GeV Higgs are the SM ones times a universal suppression factor $\cos\theta_{h,\phi}$, where $\theta_{h,\phi}$ is an unknown mixing angle between the singlet and the doublet scalars.
The measured signal strength of  $H\ra \tau\tau$, $0.98\pm0.18$\cite{CMS:htautau} and $1.09
\substack{+0.18 \\ -0.17} (stat)\substack{+0.27 \\ -0.22} (syst)\substack{+0.16 \\ -0.11} (theory)$\cite{ATLAS:htautau},  gives a relatively weak bound roughly $\sin^2\theta_{h,\phi}<0.4$ at 1 $\sigma$ if two measurements are naively combined quadratically. For $H\ra \mu\mu$, only upper bounds, $<2.8(2.92)$ form ATLAS(CMS) at 95\%C.L.\cite{LHC:hmumu}, are available. Currently, there is no constraint on the coupling between the 125GeV Higgs and the electron.

\section{ SM Gauge Interactions}
 We now return to the general case of ${\cal M}^c$ and denote
 the mass(flavor) eigenstates by $\tilde{e}(=(e,\mu,\tau,E_-,E_+))$.
  In the mass basis, the SM gauge interactions become
\beqa
&&-i \sum_{a=1}^4\sum_{i,j} \overline{\tilde{e}_i} (U_{L,ai})^\dagger \gamma^\mu \hat{L} \left[\frac{g_2}{c_W} g_L Z_\mu - e P_\mu\right]   U_{L,aj}\tilde{e}_j %\nonr\\
-i \sum_{i,j} \overline{\tilde{e}_i} (U_{R,5i})^\dagger \gamma^\mu \hat{R} \left[\frac{g_2}{c_W} g_L Z_\mu - e P_\mu\right]   U_{R,5j}\tilde{e}_j\nonr\\
&&-i \sum_{a=1}^4\sum_{i,j} \overline{\tilde{e}_i} (U_{R,ai})^\dagger \gamma^\mu \hat{R} \left[\frac{g_2}{c_W} g_R Z_\mu - e P_\mu\right]   U_{R,aj}\tilde{e}_j%\nonr\\
-i \sum_{i,j} \overline{\tilde{e}_i} (U_{L,5i})^\dagger \gamma^\mu \hat{L} \left[\frac{g_2}{c_W} g_R Z_\mu - e P_\mu\right]   U_{L,5j}\tilde{e}_j\nonr\\
&&+ H.c.\,,
\eeqa
where $P$ stands for the photon field, $c_W$ is the weak mixing, $\hat{L}/\hat{R}$ are the chirality projections, and $g_{L/R}=T_3-Q s_W^2$.
It is easy to see that the QED part is flavor diagonal in the mass basis. Since $L_2$ and $E_2$ have different chiralities comparing to their SM counterparts,  in addition to the SM neutral current(NC) and charged current(CC) interactions, one also has the following extra interactions given by
\beq
\frac{g_2}{2c_W}\left[ \overline{\tilde{e}_i}\gamma^\mu\left(g^V_{ij}-g^A_{ij}\gamma_5\right)\tilde{e}_j
-\overline{\tilde{\nu}_i}\gamma^\mu\left(g^V_{ij}-g^A_{ij}\gamma_5\right)\tilde{\nu}_j \right] Z_\mu
+\frac{g_2}{\sqrt{2}} \overline{\tilde{\nu}_i}\gamma^\mu \left(-g^V_{ij}+g^A_{ij}\gamma_5\right)\tilde{e}_j W^+_\mu +H.c.
\label{eq:new_NC_CC}
\eeq
where
\beq
g^V_{ij}\equiv \frac{1}{2}\left[(U_L^\dagger)_{i5}(U_L)_{5j}-(U_R^\dagger)_{i5}(U_R)_{5j} \right]\,,\;
g^A_{ij}\equiv \frac{1}{2}\left[(U_L^\dagger)_{i5}(U_L)_{5j}+(U_R^\dagger)_{i5}(U_R)_{5j} \right]\,.
\eeq
In general, the extra gauge interactions are flavor non-diagonal. Also, the  additional CC part of Eq.(\ref{eq:new_NC_CC}) deviates from the standard $(V-A)$ structure at low energies and  can be searched for.
The current experimental limit is roughly
\beq
|g^A_{aa}-g^V_{aa}|\lesssim 0.11\,,
\eeq
derived from  the right-handed $W_R$ boson mass limit, $M_{W_R}\gtrsim 0.7$TeV\cite{PDG}(if assuming the coupling strength equals to $g_2$).
Also, the NC part of Eq.(\ref{eq:new_NC_CC}) can induce tree-level flavor changing $Z\ra e\mu$ decay. The current bound $B(Z\ra e\mu)<7.5\times 10^{-7}$\cite{Z_emu} sets a more stringent limit that
\beq
(g^A_{12})^2+(g^V_{12})^2 < 1.4\times 10^{-6}\,,
\eeq
or roughly, $|g^A_{12}|,|g^V_{12}| \lesssim 10^{-3}$.
The above experimental limits indicate that the charged lepton mass matrix is not arbitrary in this model.

For a symmetric mass matrix as given in Eq.(\ref{eq:SCLM}), the left- and right-handed rotations
are the same. For Eq.(\ref{eq:SUR}), $U_{i5}=0$ for $i=1,2,3$  and flavor changing NC for the
SM leptons are eliminated. The low energy CC is also of the $V-A$ form. However, it predicts
flavor changing NC decays and non-standard CC reactions for the exotic leptons.

\section{Collider Phenomenology}

 The existence of a $Z_\ell$ is a robust prediction of a broken gauged $\Uel$, and its mass $M_X$ is a free
 parameter. It has vector couplings to the charged leptons in the initial gauge basis.
In the mass basis, this coupling matrix for EYL is given by\footnote{ In general, the left- and right-handed mass diagonalizing matrices are different. Then $Q_\ell^{L/R}=U^\dagger_{L/R}\cdot Q_{0\ell}\cdot U_{L/R}\,$. Charged lepton flavor violation (CLFV) couplings are expected.}

\beq
Q_l\equiv U^T \cdot Q_{0\ell}\cdot U=U^T\cdot \left(
           \begin{array}{ccccc}
            1&&&&\\
               &1&&\text{\Huge 0}&\\
                  &&-1&&\\
                     &\text{\Huge 0}&&-1&\\
                        &&&&0\\
           \end{array}
         \right)
 \cdot U =\left(
           \begin{array}{ccccc}
            1&0&0&\multicolumn{2}{c}{}\\
            0&-1&0&\multicolumn {2}{c}{\raisebox{1.5ex}[0pt]{\text{\huge 0}}}\\
            0&0&0&\frac{1}{\sqrt{2}}&\frac{-1}{\sqrt{2}}\\
            \multicolumn{2}{c}{}&\frac{1}{\sqrt{2}}&\multicolumn{2}{c}{}\\
            \multicolumn{2}{c}{\raisebox{1.5ex}[0pt]{\text{\huge 0}}}&-\frac{1}{\sqrt{2}}&
            \multicolumn{2}{c}{\raisebox{1.5ex}[0pt]{\text{\huge 0}}} \\
           \end{array}
         \right)
\eeq

Note that at the leading order, there is no tree-level $\tau^+ \mhyphen\tau^- \mhyphen Z_\ell$ coupling.
Also, there are no $\mu\mhyphen e \mhyphen Z_\ell$ couplings.

Assuming that $M_X\gg v$, the following 4-lepton operators will be generated by integrating out $Z_\ell$,
\beq
 \frac{g_{\ell}^2}{ M_X^2} \left(\frac{1}{2} \bar{e}\gamma^\mu e \bar{e}\gamma_\mu e - \bar{e}\gamma^\mu e \bar{\mu}\gamma_\mu \mu \right)\,.
\eeq
When $\sqrt{s}<M_X$, the contribution from $Z_\ell$ mediated processes  are destructive and constructive
 relative to the SM one for $e^+e^-\ra e^+e^-$ and $ e^+e^-\ra \mu^+\mu^- $, respectively.
From the corresponding 95\%C.L. limits given by LEP2\cite{LEP2},  the most constraining bound is
\beq
\frac{g_\ell}{M_X}<\frac{1}{5.33\mbox{TeV}}\,,
\label{eq:4lepton_B}
\eeq
derived from that $\Lambda^+_{\mu\mu}>18.9$ TeV. In other word, $v_L>7.54$TeV and
\beq
M_X> 1.67\times \left(\frac{g_\ell}{e}\right) \mbox{TeV}\,.
\eeq
Therefore, this model cannot accommodate  the observed $\Delta a_\mu$ anomaly by $Z_l$ alone.

\begin{figure}[htb]
 \centering
 \includegraphics[width=0.8\textwidth]{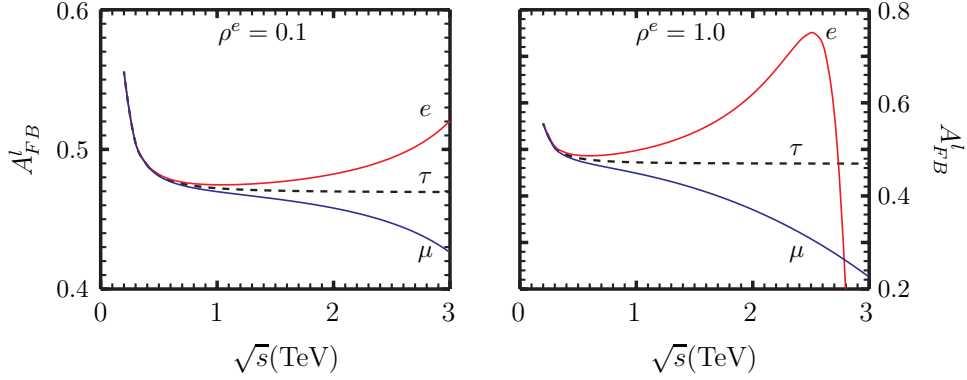}
 \caption{ $A_{FB}$  v.s. $\sqrt{s}$(GeV)  in our model.
 We take $M_X=4$TeV, $\rho^e=0.1(1.0)$ for the left(right) panel.
 The upper(lower) curve is for $A_{FB}^{e(\mu)}$, and the middle one is for $A_{FB}^\tau$.}
\label{fig:AFB}
 \end{figure}

On the other hand, the collider signals are more promising.
The decay signal of $Z_\ell\ra e^+e^-,\mu^+\mu^-$ will be clean and unambiguous if the on-shell $Z_\ell$ can be produced at the future colliders. However,
the flavor non-universal $Z_\ell$ couplings can be tested at the near-future $e^+e^-$ colliders even the c.m. energy, $\sqrt{s}$, is below $M_X$.
The contribution from $Z_\ell$ will interfere with the SM ones mediated by $Z,\gm$.
For $M_Z<\sqrt{s} <M_X$, the differential cross section for $e^+e^-\ra f\bar{f}$  is given by
\beqa
\label{eq:eeff}
\frac{d \sigma^f}{d x} &=& \left. N_c^f\frac{\pi \alpha^2}{2 s}\times \right\{  (D^f_{\gamma \ell})^2(1+x^2)\nonr\\
&&%\left.
+\frac{D_Z^2}{4(s_W c_W)^4}\left[[(g^e)_L^2+(g^e_R)^2][(g^f)_L^2+(g^f_R)^2] (1+x^2)%\nonr\\
%&&\;\;\;\; \left. \phantom{ \frac{11111}{22222} }
+2[(g^e)_L^2-(g^e_R)^2][(g^f_L)^2-(g^f)_R^2] x \right] \nonr \\
&&\left.+\frac{D^f_{\gamma \ell}D_Z}{2 (s_W c_W)^2}  \left[(g^e_L+g^e_R)(g^f_L+g^f_R) (1+x^2)+ 2(g^e_L-g^e_R)(g^f_L-g^f_R)x\right] \right\}\,,
\eeqa
where $x=\cos\theta$, $\theta$ is the scattering angle between particle $f$ and the incident $e^-$, $N_c^f$ is the color factor of $f$, $c_W(s_W)$ is the cosine(sine) of the weak mixing angle,
and $g^f_L=T_3(f)-Q_f s_W^2$ and $g^f_R = -Q_f s_W^2$ are the SM Z-fermion couplings.
The flavor-dependent dimensionless gauge boson propagator factors are also introduced\footnote{The widths, $\Gamma_Z$ and $\Gamma_X$, can be trivially put back when $\sqrt{s}$ is close to either of the two poles.},
\beq
D^f_{\gamma \ell}= -q_f + \frac{\rho^f  }{ 1-M_X^2/s }\,,\,\, D_Z= \frac{1}{1- M_Z^2/s }\,,
\eeq
where $q_f$ is the electric charge of $f$, $\rho^e= (g_{\ell}/e)^2$, $\rho^\mu= -(g_{\ell}/e)^2$, and $\rho^f=0$ for $f\neq e,\mu$.
The photon and $Z_\ell$ exchange are combined together since both have vector couplings to $e$ and $\mu$.
The forward-backward asymmetry,
\beq
A_{FB}^f = {\sigma_F-\sigma_B \over \sigma_F+\sigma_B}\,,\,\mbox{where}\;
\sigma_F=\int^1_0 dx \frac{d \sigma^f}{dx}\,,\;
\sigma_B=\int^0_{-1} dx \frac{d \sigma^f}{dx}\,,
\eeq
can be easily read from Eq.(\ref{eq:eeff}).
One example is shown in Fig.\ref{fig:AFB} for a $4$TeV $Z_\ell$ with $\rho^e=1.0, 0.1$.
Moreover, if $|\rho^e|=0.1(0.01)$,  the required c.m. energy is roughly $\sqrt{s}\sim 0.62(0.93) M_X$ for a clear $10$\% difference, $A_{FB}^e/A_{FB}^\mu=1.10$, to be observed.

For a more general mass matrix, flavor changing $Z_\ell$ couplings are expected; hence tree-level CLFV processes are possible.
For example, the rare $\mu\ra 3e$ process can be generated by exchanging a $Z_\ell$ .
Following \cite{LFV_exD}, one has
%\beq
%g_3=g_4=g_5=g_6= { \sqrt{2} g_{\ell}^2 Q_l^{ee}Q_l^{e\mu}\over 8 G_F M_X^2},
%\eeq and
\beq
Br(\mu\ra 3 e)= \frac{3}{4 G_F^2} \left(\frac{g_\ell}{M_X}\right)^4 |Q_l^{ee} Q_l^{\mu e}|^2\,.
\eeq
Assuming that $|Q_l^{ee}|\sim 1$, from Eq.(\ref{eq:4lepton_B}) and that $Br(\mu\ra 3e)< 10^{-12}$\cite{PDG}, we get $|Q_l^{\mu e}|\lesssim 4\times 10^{-4}$.
A similar analysis for CLFV three-lepton tau decays with $Br(\tau\ra 3l)\lesssim 10^{-8}$\cite{PDG} give weaker bounds that $|Q_l^{\mu \tau}|,|Q_l^{e \tau}|\lesssim 10^{-1}$.
Moreover, for $\sqrt{s}<M_X$,  the flavor violating branching fraction
 at the $e^+e^-$ colliders can be estimated to be
\beq
B_{ij}\equiv {\sigma(e^+e^-\ra l_i l_j) \over \sigma(e^+e^-\ra \mu^+\mu^-)} \simeq \frac{g_{\ell}^4}{e^4} {|Q_l^{ee} Q_l^{ij}|^2 \over (1-M_X^2/s)^2 }\,,\;\mbox{where}\; i\neq j\,.
\eeq
For example, if  $\sqrt{s}=1$TeV and $M_X=4$TeV, $B_{\mu e}\lesssim 10^{-8}$ and $B_{\tau e, \tau \mu}\lesssim 10^{-4}$ can be derived.
Therefore, if the CLVF $\tau$ decay branching ratios are close to the current bounds, the $\sqrt{s}$-dependent  $e^+e^-\ra \mu\tau, \tau e$ could be observed in the future $e^+e^-$ colliders with an integrated luminosity $\sim ab^{-1}$.

This anomaly-free arrangement requires only one-third of exotic fermions compare to the solution studied in \cite{CNL1,CNL2}.
 Therefore, the oblique parameter $\Delta S$ and $\Delta T$\cite{Peskin} contraints,
 \beq
\triangle T=\frac{1}{16\pi s_w^2 }\sum_{i=1,2}\frac{M_{E_i}^2}{M_W^2}\left(1+x_i+\frac{2x_i}{1-x_i}\ln x_i\right)\,,\,
\triangle S= \frac{1}{6\pi}\left[2 + \ln (x_1 x_2)\right]\,,
\eeq
where $x_i = M_{N_i}^2/M_{E_i}^2$, are much weaker than in \cite{CNL1,CNL2}.  Using the experimental limit of $\triangle S_{\mathrm{exp}}<0.25$ \cite{PDG}, it can be seen that even for degenerate exotic leptons they are within experimental bound. However, for exotic leptons with masses around $\sim 0.5(1.0)$TeV, $\Delta T_{\mathrm{exp}} <.32$ will require that the mass splitting between the isodoublet components have to be less than $20(10)$\%. Since the limits on the exotic charged particle mass from the direct search are around $>100$ GeV\cite{PDG}, it is expected that the charge neutral components acquire masses $\gtrsim 100$ GeV as well.

At the LHC, $Z_\ell$ can be produced via the radiative Drell-Yan process, $pp\ra e^+e^-Z_\ell$\cite{CNL1}. For the EYL scenario, the $Z_\ell$ does not decay into $\tau^+\tau^-$. The signal will be an $e^+e^-$ or a $\mu^+\mu^-$ pair with the invariant mass peaking at $M_X$.
 Neither signals will have jet activities. However, limited by the contact interaction,  the lepton-number breaking scale can only be modestly probed up to $\sim 0.5(1)$ TeV at the LHC13(30) if $S/\sqrt{B}=3$ is required as detailed in \cite{CNL1}.
Similarly, the heavy leptons can be pair produced at the LHC via the SM Drell-Yan process. Note that their production cross sections, $\sim {\cal O}(1-100 fb)$ if they are lighter than $500$GeV, are independent of $g_\ell$ and $M_X$.

\section{Discussion and Summary}
In this work, a novel arrangement to promote the approximate lepton-number conservation in the SM to an anomaly-free gauged $U(1)_\ell$ theory is presented. We have discussed the case that two out of the three SM lepton generations have the opposite $U(1)_\ell$ charges, $\zeta_1=-\zeta_2$, as in\cite{HJLV,Fo}, and the remaining one  with $U(1)_\ell$ charge $\zeta_3$ has its anomalies canceled  with four exotic leptons, $L_{1,2}$ and $E_{1,2}$ (see Table I), as introduced in \cite{CNL1,CNL2}.
 Moreover, we focus in this paper the interesting case that $|\zeta_1|=\zeta_3$ so that nontrivial generation-crossing Yukawa mixings are allowed.
One singlet scalar is added to make the two exotic charged lepton heavier than the electroweak scale and to break $U(1)_\ell$ spontaneously.  To the best of our knowledge, this solution requires the least number of new degrees of freedom to solve the anomalies for all three generations.

The resulting charged lepton masses and the SM gauge interactions have been carefully studied.
The anomaly-free particle content results in new, in general flavor-changing, SM NC and CC interactions.
The current experimental constraint on the flavor changing weak interactions suggests that the charged lepton mass matrix cannot be arbitrary.
As an illustration,  we have studied a simplified limit which satisfies the above mentioned experimental bounds, in which the Yukawa couplings are universal, and the charged lepton mass matrix is symmetric.
We have found that this model naturally predicts two out of the three SM charged leptons acquire masses much below, and the other one around,  the electroweak scale.  This delightful consequence encourages one to entertain
 the possibility that the lepton charges for the three SM generations need not be the same.

 A comprehensive discussion on the neutrino mass generation is beyond the scope of this paper.
Unlike the charged lepton masses generation, which stems from the SSB of the SM electroweak and $U(1)_\ell$, the light neutrino masses, $m_\nu$, require more model building. In a nutshell, one can either add a pair of vectorlike singlets $(1,0,\pm1)$ as in type-I seesaw, or a triplet scalar as in \cite{CNL2} for tree level $m_\nu$. Both of these scenarios require fine tuning of Yukawa couplings and/or triplet VEV. It can also be radiative generated by adding a set of scalars\footnote{ At 1-loop, two doublets with $(2,1/2,\pm2)$, a singlet with $(1,0,2)$, and a charged singlet with $(1-1,0)$ are needed for a realistic
mass matrix.} similar to \cite{CNL1}.

The phenomenology of this model has mostly to do with the exotic degrees of freedom, similar to the discussion in \cite{CNL1}. However, $Z_\ell$ phenomenology differs from the previous one since now the leptons have distinctive $U(1)_\ell$ charges. A robust prediction is that $e,\mu, \tau$ have different forward-backward-asymmetries at the $e^+e^-$ colliders and can be searched for. Moreover, the flavor changing
processes $e^+e^-\ra \tau\mu, \tau e$ can be anticipated at the $e^+e^-$ collider, with $\sqrt{s}\sim $TeV and an integrated luminosity $\sim ab^{-1}$, if the branching ratios $Br(\tau\ra 3 l)$ are not too much smaller than the current limits.

\begin{acknowledgments}
We would like to thank Dr. D. McKeen for reminding us that $\ell_\mu -\ell_\tau$ is anomaly-free.
 WFC is supported by the Taiwan Ministry of Science and Technology under
Grant No. 106-2112-M-007-009-MY3.
TRIUMF receives federal funding via a contribution agreement with the National Research Council of Canada and the Natural Science and Engineering Research
Council of Canada.
\end{acknowledgments}

\end{document}